\documentclass[3p,times,twocolumn]{elsarticle}

\usepackage{ecrc}

\volume{00}

\firstpage{1}

\journalname{Nuclear Physics B Proceedings Supplement}

\runauth{N. Sartore}

\jid{nuphbp}

\jnltitlelogo{Nuclear Physics B Proceedings Supplement}

\usepackage{amssymb}

\usepackage[figuresright]{rotating}
\usepackage{graphicx}
\usepackage{graphics}
\usepackage{txfonts}
\usepackage{eps2pdf}

\begin{document}

\begin{frontmatter}




\title{Science with the ASTRI prototype}

\author[label1]{Nicola Sartore}
\address[label1]{INAF - Istituto di Astrofisica Spaziale e Fisica Cosmica - Via E. Bassini 15, Milano, Italy}

\author{on behalf of the ASTRI Collaboration}

\address{sartore@iasf-milano.inaf.it}

\begin{abstract}
ASTRI (Astrofisica a Specchi con Tecnologia Replicante Italiana) is a ``Flagship Project'' financed by the Italian Ministry of Instruction, University and Research 
and led by the Italian National Institute of Astrophysics. It represents the Italian proposal for the development of the Small Size Telescope system of the 
Cherenkov Telescope Array, the next generation observatory for Very High Energy gamma-rays (20 GeV - 100 TeV). The ASTRI  end-to-end prototype will be installed at 
Serra La Nave (Catania, Italy) and it will see the first light at the beginning of 2014. 
We describe the expected performance of the prototype on few selected test cases of the northern emisphere. The aim of the prototype is to probe the technological 
solutions and the nominal performance of the various telescope's subsystems.
\end{abstract}


\end{frontmatter}


\section{INTRODUCTION}\label{sect-intro}
The advent of the Imaging Atmospheric Cherenkov Telescopes (IACTs) has greatly improved our knowledge of the Very High Energy (VHE) sky, allowing for the 
detection of VHE counterparts of known sources as well as to discover previously unknown types of VHE emitters.
The number of VHE sources is currently well beyond 100 and it will grow further with the planned Cherenkov Telescope Array (CTA, \cite{actis}), 
which is expected to reach a sensitivity 10 times lower than that achieved by present-day IACTs over a wider energy range ($\sim 10$ GeV - 100 TeV).
The detection of photons in the highest energy window of CTA ($\sim10 - 100$ TeV) will be possible thanks to a large array of tens of Small Size Telescopes (SST), 
with a diameter of the mirror of about 4 meters.

In this context, the Italian National Institute of Astrophysics is leading the ASTRI (``Astrofisica a Specchi con Tecnologia Replicante Italiana'') project, 
a flagship project of the Italian Ministry of the Education, University and Research, aimed at the development of an end-to-end prototype of the 
SST-2M (i.e. based on the dual mirror technology, see Section \ref{sect-sst}) and the replica technology necessary for series production of the SSTs.
The ASTRI prototype will be installed at Serra la Nave, 1735m a.s.l. on the Mount Etna (Sicily), at the INAF ``M. G. Fracastoro'' observing station.
The main structure is expected to be on-site by October 2013. After the integration of the different subsystems and the commissioning procedures, 
the start of the science verification phase is planned to begin in February 2014.

\section{DESCRIPTION OF THE SST-2M PROTOTYPE}\label{sect-sst}
\begin{figure}[tb]
\centering
\includegraphics[width=105mm,height=80mm,clip,trim=0mm 0mm 0mm 110mm]{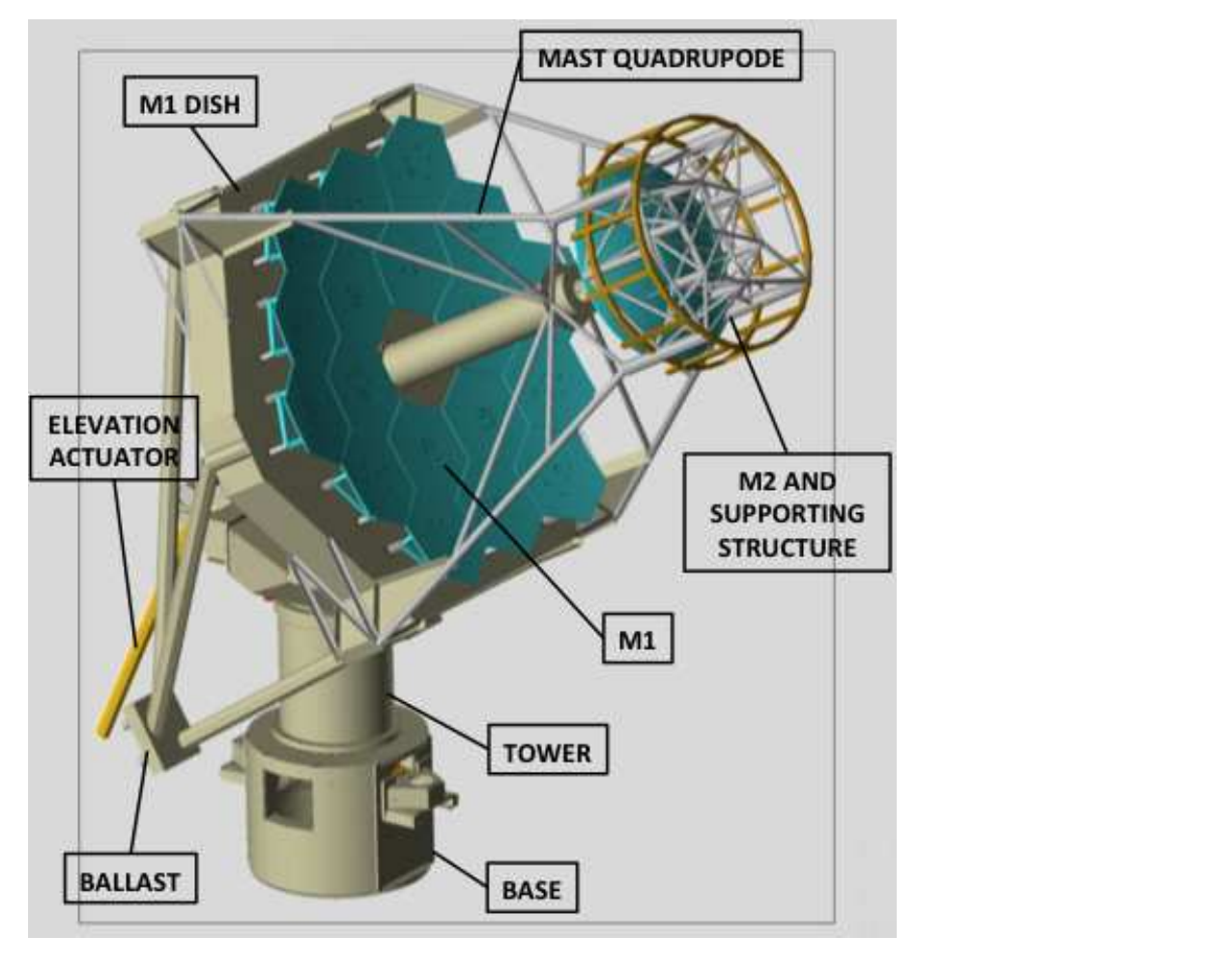}
\caption{Scheme of the telescope structure and optical design of the SST-2M prototype. Courtesy of R. Canestrari}
\label{fig-sst}
\end{figure}

The SST-2M prototype will incorporate a number of innovative features, some of which represent a technological challenge.
The telescope will have a dual mirror Schwarzschild-Couder configuration (Figure \ref{fig-sst}), the first time among IACTs. 
The primary mirror is aspherical and is segmented into 18 elements, for a total diameter of 4.3 meters \cite{canestrari}.
The secondary, monolithic mirror has a diameter of 1.8 meters and is placed at 3 meters from the primary. The field of view (FoV) is 9.6 degrees and 
the focal ratio is 0.5. The compactness of the design guarantees a reduction in the weight and necessary robustness of the telescope structure, 
thus reducing the production costs.

The camera uses Hamamatsu Silicon Photomultipliers type S11828-3344M \cite{hamamatsu}. To cover the whole FoV, we used a modular setup for the camera, 
where the basic physical unit is made of 4x4 pixel ($\rm 3mm\times3mm$ each), whereas the basic logical unit is made of 2x2 pixel, equivalent to a FoV of 
0.17 degrees. The modular setup assures an easy and rapid maintenance of single parts of the camera.

\section{SELECTION OF TARGET CANDIDATES}\label{sect-obs}

\begin{figure}
\flushleft
\includegraphics[width=70mm,height=80mm,angle=90,clip,trim=10mm 5mm 10mm 25mm]{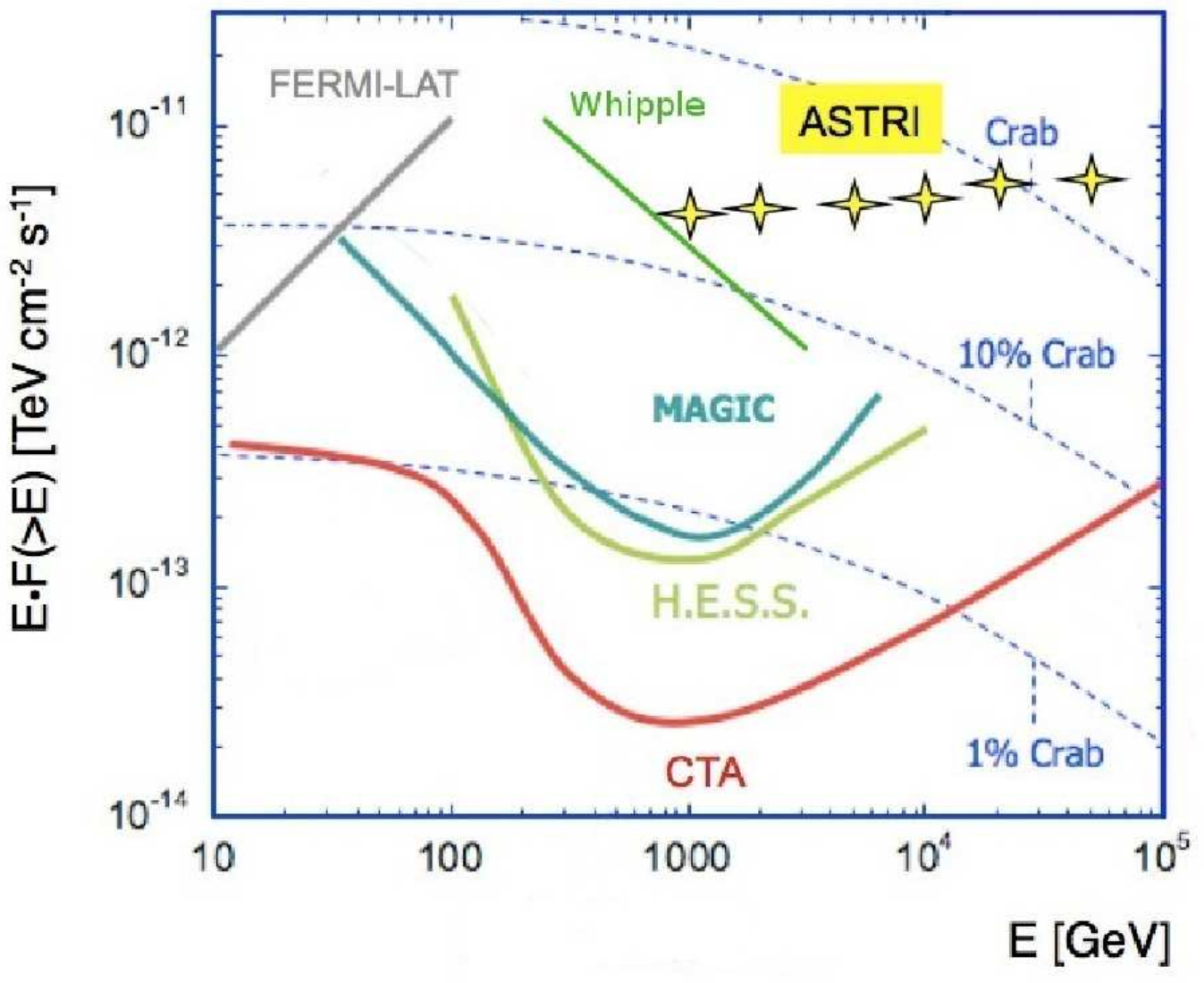}
\caption{Comparison of the integral sensitivity of ASTRI and CTA with those of current HE/VHE instruments.}
\label{fig-sens}
\end{figure}

Preliminary estimates \cite{vallania} of the telescope response to gamma and cosmic rays suggest the performance of the SST-2M 
are worse than those achieved by the present-day IACTs, and comparable at best with that of the {\it Whipple} telescope (Figure \ref{fig-sens}). 
Thus, we used the information available on the online catalogue of TeV sources\footnote{\it http://tevcat.uchicago.edu/.}, 
in order to select an essay of possible candidates among the brightest sources. 
Due to its location in the northern emisphere, the sky accessible to the SST-2M prototype is approximately the same as the VERITAS and MAGIC telescopes.
We selected only sources of the northern sky for which the flux reported in the TeV catalogue is a substantial fraction of the Crab Nebula flux.
Owing to the low energy resolution expected for the SST-2M prototype \cite{vallania}, we simulate observations of the integral flux only, 
in the $1-100$ TeV energy range.

To constrain the observability of the selected targets with the SST-2M we use the following approach: we perform a convolution of the source spectrum with the 
effective area of the SST-2M inferred from analytic calculations, in order to obtain the rate of observed events.
We then compare the signal from the source with that of the cosmic ray background \cite{antoni04}, assuming that our gamma/hadron discrimination 
procedure retains 50\% of the gamma-rays and discards 90\% of the cosmic rays \cite{vallania}. Additional effects like variations of the observing conditions 
(nightly variations of the source's zenith angle due to the sideral motion of the Earth and of the night sky background) are not taken into account.
Hereafter, we briefly describe the results of the simulated observation runs obtained for each source, listing first the Galactic sources, 
followed by the extragalactic ones.

{\it Crab Nebula}. This pulsar wind nebula is the brightest Galactic source at TeV energies.
For our evaluation we use the log-parabolic spectrum reported in \cite{aleksic12}. We obtain a rate of $\sim100$ excess photons/hour, which translates in a 
$\sim23 \sigma$ detection in 50 h of observations. A $5\sigma$ detection could be obtained in $\sim2.5$ hours.

{\it LS I $+61^\circ\,303'$}. The system is a high-mass X-ray binary composed by a Be star and a compact source of an unknown type, orbiting each other in highly 
eccentric orbit ($e=0.72$) with a period of $\sim26$ days. The emission is observed at almost all wavelengths. In particular, modulated TeV emission 
period has been reported by both MAGIC \cite{albert08a} and VERITAS \cite{acciari08} at fluxes of about 10\%-20\% of the Crab nebula (Crab Units, C.U. hereafter). 
However, the modulation is not properly periodic, and variations between different cycles have been reported as well.
On the other hand, the {\it Whipple} team reported an upper limit of 0.184 C.U. in $\sim5$ hours of observations \cite{smith07}.
We expected a similar result for the SST-2M prototype. Assuming the high-state spectrum reported by \cite{acciari08}, our calculations suggest that the observation
time needed for $5\sigma$ detection would be more than 200 hours. Thus, this source is far from being an ideal target for the SST-2M prototype.

{\it Cygnus Region}. This is one of the brightest sky regions, containing many TeV point-like and extended sources, and it overlaps with the Cygnus OB1 association 
of young stars. Hence, the observed emission is likely the superposition of individual sources, like e.g. pulsar wind nebulae, supernova remnants and 
VHE-emitting  high mass binaries. 
The MILAGRO collaboration reported emission from 1 to 200 TeV with an integrated flux of $\sim0.8$ C.U. for the source MGRO J2019+37
and of $\sim0.4$ C.U. for the source MGRO J2021+41 \cite{abdo12}, which in principle make this region worth of consideration. 
However, the results of the ARGO-YBJ collaboration are consistent with those of MILAGRO only for J2021+41, while for J2019+37 they reports no excess 
at $3\sigma$ level \cite{bartoli12}. Furthermore, the MAGIC and VERITAS telescopes reported fluxes significantly lower \cite{albert08b,aliu11} 
than those reported by MILAGRO and ARGO-YBJ. 
This discrepancy may be indicative of a variable emission or could arise from the different angular resolution among the various instruments, 
which imply different background estimation and subtraction, and thus significance, of their results. Due to the low flux level measured by more sensitive IACTs, 
we deem this not as a viable target for the SST-2M prototype.

{\it Markarian 421}. This $z=0.031$ redshift blazar exhibited strong TeV variability during the years, with fluxes going from below 1 up to $\sim10$ C.U.
Hence, for this source we adopt many different spectral models \cite{acciari11}, in order to constrain its observability at different flaring states.
Photon rates span from about 40 to about 600 photons per hour, and imply integration times from 14 to $\sim0.1$ hours, assuming a $5\sigma$ detection.
The possibility of high-significance detection in less than one night of observation could allow short time-scale variability studies, eventually in conjunction 
with other instruments observing at different wavelengths.

{\it Markarian 501}. Similarly to Mrk 421, this blazar showed strong flaring activity at ``super-Crab'' flux levels, 
as reported by the {\it Whipple} ($\sim1.6$ C.U. \cite{catanese97}) and CAT experiments ($\sim10$ C.U., \cite{djannati-atai99}) and, recently, 
also by the ARGO-YBJ experiment \cite{bartoli12} ($\sim2.0$ C.U. during the 2011 flare).
In the low state the reported flux is $\sim0.3$ C.U. Thus, the considerations made for Mrk 421 hold also in this case. 
Adopting the spectral models proposed by \cite{djannati-atai99}, the rate of excess photon lie in the 40-3000 photons/hour range, 
implying observation times from 15 hours to few minutes, depending on the flaring state.

{\it M87}. This radio galaxy is the first known VHE sources of its type. As in the cases of the two Markarian blazars mentioned above, its emission is highly  
variable \cite{tescaro09}. However, even in the flaring state the flux level is barely $\sim15$ percent of that of the Crab nebula. For the spectral model proposed 
by \cite{tescaro09}, the excess photon rate is $\sim20$ photons per hour, and $5\sigma$ detection is barely achievable in at least 50 hours of integration time.
This makes M87 a likely target of opportunity for the SST-2M prototype only during periods of strong flaring activity.

It is thus clear that the scientific capabilities of the SST-2M prototype are limited by its far from competitive sensitivity.
Even so, the most easily detectable sources, that is the Crab nebula, the blazars Mrk 421 and 501, and the Cygnus region, could be suitable targets for a 
dedicated monitoring campaign that would be hampered mainly by the visibility of the targets themselves. Furthermore, the mutual overlap between the best visibility 
period of each source is negligible, allowing in principle a prolonged monitoring lasting some months \cite{bonnoli}.

\section{CONCLUSIONS}\label{sect-end}
As we already pointed out, the main objective for the ASTRI project is the realization and evaluation of a prototype of the SST-2M and its subsystems under field 
conditions. From a scientific point of view, however, its performance are worse than those of the other IACT facilities in operation and the single telescope 
analysis procedures are very different from those in stereo mode that will be used in CTA.
Nevertheless, the prototype could work as a dedicated instrument for the monitoring of the brightest TeV sources. In particular, the blazars Mk 421 and 501 are known 
for recurrent flaring activity at fluxes up to 5-10 Crab units. This will allow variability studies within a single night of observation.

On the other hand, the recently proposed SSTs mini-array, to be placed at the CTA southern site, could represent a significant improvement in terms of performance 
over the SST-2M prototype. Preliminary results of MC simulations suggest that the sensitivity of an array of 7 SSTs could be a factor 1.5 w.r.t. that of HESS 
\cite{dipierro}, thus enabling the first CTA science activities to be carried out \cite{vercellone}.

\subsection*{Acknowledgments}
This research is carried out in the framework of the MIUR-INAF  ASTRI flagship program.




\nocite{*}



\end{document}